\documentclass[useAMS,usenatbib,usegraphicx]{mn2e}
\usepackage{amsmath}

\title[SDSS J0926+3624 Orbital Expansion]{Direct Detection of SDSS J0926+3624 Orbital Expansion with ARCONS}
\author[P. Szypryt et al.]{P. Szypryt$^{1}$\thanks{E-mail: pszypryt@physics.ucsb.edu}, G.E. Duggan$^{1,2}$, B.A. Mazin$^{1}$, S.R. Meeker$^{1}$, M.J. Strader$^{1}$, 
\newauthor J.C. van Eyken$^{1}$, D. Marsden$^{1}$, K. O'Brien$^{3}$, A.B. Walter$^{1}$, G. Ulbricht$^{1}$, 
\newauthor T.A. Prince$^{2}$, C. Stoughton$^{4}$, and B. Bumble$^{5}$\\\\
$^{1}$Department of Physics, University of California, Santa Barbara, CA 93106, USA\\
$^{2}$Division of Physics, Mathematics, and Astronomy, California Institute of Technology, Pasadena, California 91125\\
$^{3}$Department of Physics, University of Oxford, Denys Wilkinson Building, Keble Road, 
Oxford, OX1 3RH, UK\\
$^{4}$Fermilab Center for Particle Astrophysics, Batavia, IL 60510, USA\\
$^{5}$NASA Jet Propulsion Laboratory, 4800 Oak Grove Drive, Pasadena, CA 91125, USA}

\begin{document}

\date{Received: }

\maketitle

\begin{abstract}
AM Canum Venaticorum (AM CVn) stars belong to a class of ultra-compact, short period binaries with spectra dominated largely by helium. SDSS J0926+3624 is of particular interest as it is the first observed eclipsing AM CVn system.  We observed SDSS J0926+3624 with the \textbf{Ar}ray \textbf{C}amera for \textbf{O}ptical to \textbf{N}ear-IR \textbf{S}pectrophotometry (ARCONS) at the Palomar 200" telescope. ARCONS uses a relatively new type of energy-resolved photon counters called Microwave Kinetic Inductance Detectors (MKIDs). ARCONS, sensitive to radiation from 350 to 1100 nm, has a time resolution of several microseconds and can measure the energy of a photon to $\sim$10$\%$. We present the light curves for these observations and examine changes in orbital period from prior observations. Using a quadratic ephemeris model, we measure a period rate of change $\dot{P} = (3.07 \pm 0.56)\times 10^{-13}$. In addition, we use the high timing resolution of ARCONS to examine the system's high frequency variations and search for possible quasi-periodic oscillations (QPOs). Finally, we use the instrument's spectral resolution to examine the light curves in various wavelength bands.  We do not find any high frequency QPOs or significant spectral variability throughout an eclipse.
\end{abstract}


\section{Introduction}
The AM Canum Venaticorum (AM CVn) stars define a class of short-period binary stars with spectra dominated largely by helium. They consist of a white dwarf primary accreting helium-rich matter from a less massive secondary, typically through an accretion disk. They appear as variable, faint blue stars. The earliest review of AM CVn systems is given by \citealt{warner95}. More recent reviews are given by \citealt{nelemans05}, \citealt{ramsay07}, and \citealt{solheim10}.

AM CVn systems are believed to start as detached binaries. After one or more common envelope events, they are brought closer together due to gravitational wave driven angular momentum loss (\citealt{kraft62}). At this point, Roche-lobe overflow (RLOF) may lead to the formation of an AM CVn star. If gravitational wave radiation is the dominant phenomenon in the AM CVn evolution, the orbital period of the binary will decrease. On the other hand, if mass transfer due to RLOF dominates the evolution, the period will hit a minimum before beginning to increase (\citealt{solheim10}).  Due to their short periods, AM CVn stars are predicted to be some of the strongest sources of gravitational wave radiation. For this reason, AM CVn stars will be among the first objects studied by proposed gravitational wave missions such as LISA (\citealt{nelemans04}).

There are three possible scenarios for the formation of an AM CVn system, each of which contains an accreting white dwarf primary and a helium-rich donor. The first scenario involves a double white dwarf system which loses angular momentum due to gravitational wave radiation, decreasing the orbital period.  When the orbit becomes close enough, the lower mass, helium-rich donor white dwarf begins to transfer mass through stable RLOF, causing the period to increase. The mass transfer rate drops at this stage (\citealt{paczynski67}). In the second case, the donor star is a low-mass non-degenerate helium star. Much like in the white dwarf donor channel, after the period passes through a minimum, it begins to increase while the mass transfer rate decreases (\citealt{iben91}). In the third and least likely case, the system forms as a regular cataclysmic variable (CV) with a highly evolved secondary star. After the secondary star transfers away much of its hydrogen to reveal its helium core, it follows a path similar to the non-degenerate helium star channel (\citealt{podsiadlowski03}).

The prototype AM CVn star was initially discovered as a blue star by \citealt{malmquist36}. AM CVn was later found to be variable on very low levels by \citealt{smak67}. Soon after, \citealt{paczynski67} explained this system as a binary whose evolutionary physics is determined by gravitational wave radiation and saw this as a testing ground for general relativity. Since then, 36 new AM CVn systems have been found. The Sloan Digital Sky Survey (SDSS) and the Palomar Transient Factor (PTF) found the majority of these systems (see Table 6.1 in \citealt{levitan13}).

A particularly interesting system, SDSS J0926+3624, was discovered by \citealt{anderson05}. It was the first eclipsing AM CVn system discovered, and only recently has a second partially eclipsing AM CVn system (PTF1 J1919+4815) been found (\citealt{levitan13}). SDSS J0926+3624 has an orbital period of 28.3 minutes, deep eclipses lasting $\sim$1.3 minutes, and a mean g'-band magnitude of $\sim$19. The magnitude is quite variable throughout the orbit due to superhumping, a phenomenon that is observed in the majority of AM CVn stars (\citealt{solheim10}). Superhumping is due to the large mass ratios causing tidal stress asymmetries. These asymmetries deform the originally circular disk into a precessing elliptical disk (\citealt{whitehurst88, warner95}). 

The eclipsing nature of SDSS J0926+3624 is especially important in that it provides precise timing information. The eclipse timing information from observations spanning just a few years can be used to determine a period change caused by mass transfer and gravitational wave radiation. This can be used as a probe to study the physics of gravitational waves and check predictions of general relativity (\citealt{solheim10}).

In 2012 we observed SDSS J0926+3624 with the \textbf{Ar}ray \textbf{C}amera for \textbf{O}ptical to \textbf{N}ear-IR \textbf{S}pectrophotometry (ARCONS; see \citealt{mazin10, mazin13}). The goal of these observations was to add to the 2006 and 2009 observations performed by \citealt{marsh07} and \citealt{copperwheat11} and to use this large timespan of data to measure the orbital period change, $\dot{P}$. The use of ARCONS to make these observations also allowed us to probe new regions of parameter space. Using the microsecond time resolution of ARCONS, we searched for quasi-periodic oscillations (QPOs) in much higher frequency space than had previously been possible. QPOs have been observed in other variable sources such as CVs and X-ray binaries (\citealt{warner08}). Finally, we used the instrument's photon energy resolution to examine the light curves of SDSS J0926+3624 in multiple bands from blue to infrared. 

\section{Instrument}
ARCONS uses a new superconducting technology called Microwave Kinetic Inductance Detectors (MKIDs; see \citealt{day03, mazin12}). MKIDS are nearly ideal photon sensors, capable of measuring the energy of a photon to within a few percent and the arrival time to a microsecond. There is no read noise or dark current.
The array used during this particular observing run contains a total of 2024 (44x46) pixels. The plate scale is 0.45 arcseconds/pixel, making the field of view roughly 20x20 arcseconds. ARCONS is sensitive to photons in the 350-1100 nm range and has an energy resolution $E/\delta E=8$ at 400 nm.

\section{Data}

SDSS J0926+3624 was observed at the Palomar 200" telescope over the course of three nights in December, 2012.  Observations took place on the nights of December 8, 10, and 11. Seeing stayed between 1 $\textendash $1.5" throughout the three nights of observation. The count rates began to rise toward the end of each night of observations due to the onset of twilight. A summarized log of the observations is shown in Table~\ref{table:observation}.
\begin{table*}
\centering
\begin{minipage}{140mm}
\caption{Observation Log}
\begin{tabular}{c c c l}
\hline\hline
Night & UTC & Eclipses & Comments \\ 
 & Start End & observed &  \\ [0.5ex]
\hline
December 8, 2012 & 12:11 13:28 & 3 & Seeing $\sim$1.5", short break for wavelength calibration \\
December 10, 2012 & 07:44 09:01 & 2 & Seeing 1 $\textendash$ 1.5", poor focus, data omitted \\
 & 12:02 13:23 & 3 & Seeing $\sim$1.5", telescope refocused, poor conditions \\
December 11, 2012 & 11:27 13:33 & 5 & Seeing 1 $\textendash$ 1.5", thin high clouds \\
\hline
\end{tabular}
\label{table:observation}
\end{minipage}
\end{table*}

Guiding was done using an SBIG STF-8300M CCD Camera, which has a field of view of $\sim$1.5 arcminutes. A guide star was tracked by using the camera in 3x3 binning mode with exposure times of 10-15 seconds, depending on observing conditions. Due to technical constraints on the cryogenic system, the instrument was mounted at Coud\'e focus. The resulting field rotation was taken into account in the guiding software.

After the observational data was read out, it was stored in HDF\footnote{http://www.hdfgroup.org/} files. From there, it was pushed through the ARCONS data reduction pipeline, as detailed in \citealt{mazin13}.  The pipeline steps used included dead pixel masking, cosmic ray cleaning, wavelength calibration and flatfield calibration.

\begin{figure*}
\centering
\begin{center}
\includegraphics[width=\linewidth]{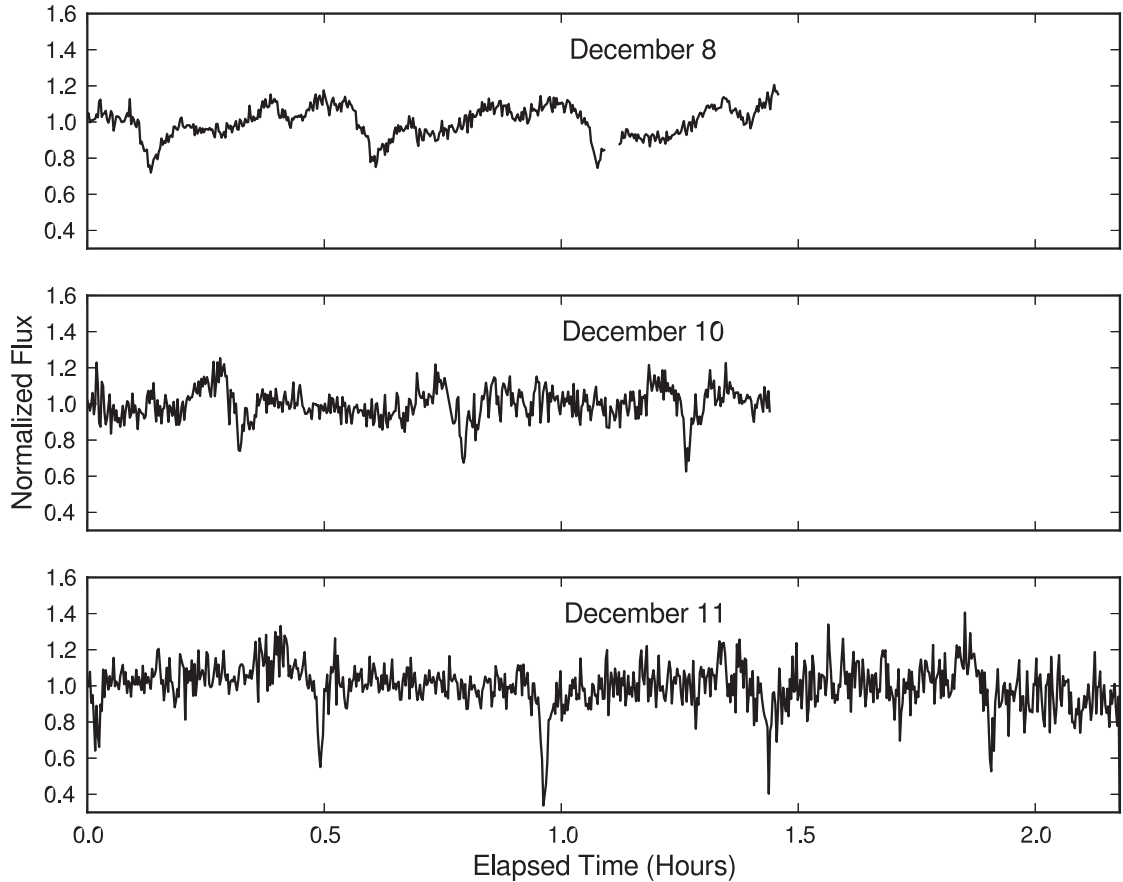}
\end{center}
\caption{The light curves of SDSS J0926+3624 from December 8, 10, and 11. The integration time is set to 10s and only photons within the 4000-5500 \AA\ range are used.  The zero point in time marks the beginning of an observation during a particular night.
} 
\label{fig:lightcurve}
\end{figure*}

Once the data went through these reduction steps, the photons were binned by wavelength and summed over a desired integration time. ARCONS continuously detects individual photons, eliminating the need for a traditional exposure time as in a CCD and allowing us to choose integration times during data processing that best fit the application. After the initial reduction, a circular two-dimensional Gaussian point spread function (PSF) was fit to each image, and the baseline of the fit was subtracted off, corresponding to removing the background sky level. The amplitude and width were used to find the flux from the object. The light curves from all three nights of observation using a 10s integration time are shown in Figure~\ref{fig:lightcurve}. In order to maximize the signal-to-noise ratio (SNR), only photons with wavelengths between 4000-5500 \AA\ were used.

\begin{figure*}
\centering
\begin{center}
\includegraphics[width=\linewidth]{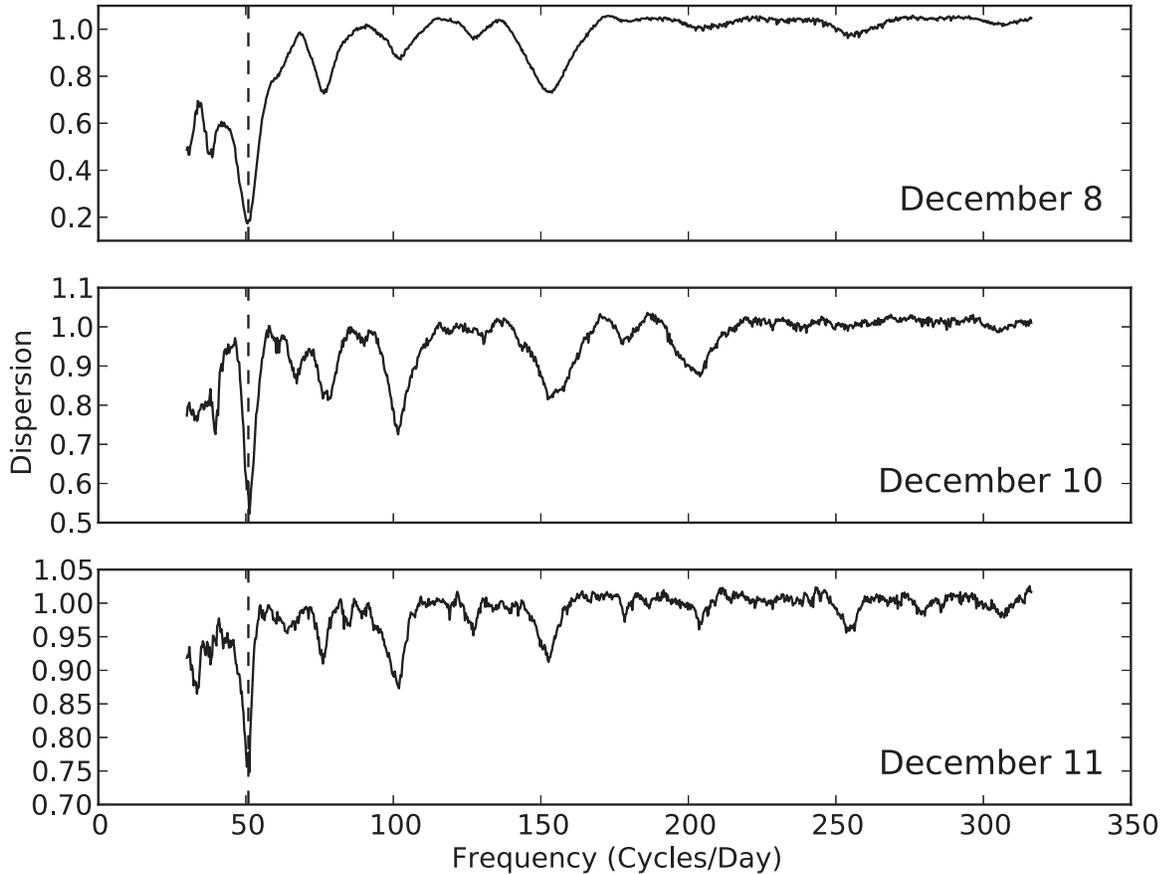}
\end{center}
\caption{Phase dispersion minimization results for the December 8, 10, and 11 data. The dashed line marks the measured eclipse frequency of $\sim$50.9 cycles/day. There is a clear dip in dispersion at the eclipse frequency and its harmonics for each night of observation
} 
\label{fig:pdm}
\end{figure*}

After the image stacks were created, we used the timing and PSF fit flux information from the light curves to perform phase dispersion minimization (PDM; see \citealt{stellingwerf78}). The results are shown in Figure~\ref{fig:pdm}. PDM was used over standard fast fourier transform (FFT) techniques due to the nonsinusoidal nature of our light curves on the observed timescale. Again, an integration time of 10s was used, and the selected photons were in the 4000-5500 \AA\ range. In each night, the dip corresponding to the eclipse frequency of $\sim$50.9 cycles per day was clearly visible. The superhump frequency, which is expected to be slightly lower than the eclipse frequency (\citealt{warner95}), could not be distinguished from the eclipse frequency. The fact that our data did not reveal clear superhumping behavior could be due to a vertical extension of the bright spot (\citealt{copperwheat11}). Observing conditions were substantially worse on December 10 and 11, and this data was not used in our higher frequency analysis.

\section{Results}

\subsection{Light Curve Analysis}

The primary focus of our light curve analysis was to precisely determine the timings of eclipses. With this we could use data from prior observations made by \citealt{copperwheat11} to better constrain the ephemeris. We started with the light curves shown in Figure~\ref{fig:lightcurve}, except we used an integration time of 3 seconds for the better quality December 8 data in order to calculate the eclipse timing more precisely.  As the observing conditions were worse on December 10 and 11, it was more difficult to perform PSF fitting photometry and a 10 second integration time was required.

\begin{figure}
\centering
\begin{center}
\includegraphics[width=\linewidth]{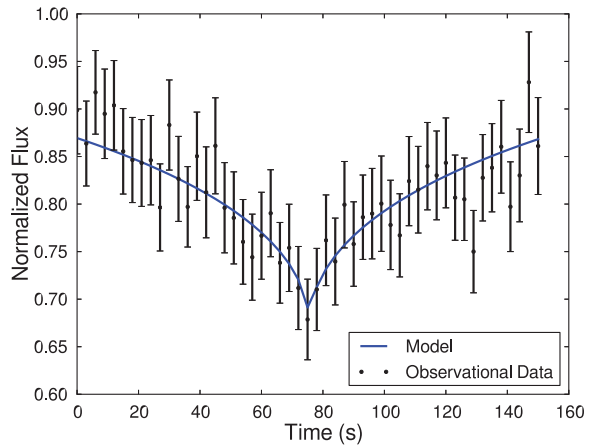}
\end{center}
\caption{Example of the model and fit used to determine the eclipse centers. Error bars in flux are calculated using the PSF fitting errors. This particular fit shows the first eclipse from the December 8 data.
}
\label{fig:exampleFit}
\end{figure}

We fit the light curves to a model containing only the white dwarf eclipse using Levenberg-Marquardt minimization. This is a reasonable model as the bright spot component is completely distinct in time from the white dwarf component, and it varies from eclipse to eclipse. We modeled the white dwarf eclipse as a limb-darkened sphere, using a square root limb-darkening law (\citealt{diaz92}). This law has the form
\begin{align}
\frac{I(\mu)}{I(1)} = 1-a_{1}(1-\sqrt{\mu})-a_{2}(1-\mu),
\end{align}

\noindent
where $\mu = \cos{\gamma}$, and $\gamma$ is the angle between the line of sight and the emergent radiation. The constants $a_{1}$ and $a_{2}$ were determined by fitting individual eclipses. We used this model to fit a time to the center of each observed eclipse. An example of the model fit for the first eclipse observed in the December 8 data is shown in Figure~\ref{fig:exampleFit}.
\begin{figure*}
\centering
\begin{center}
\includegraphics[width=\textwidth]{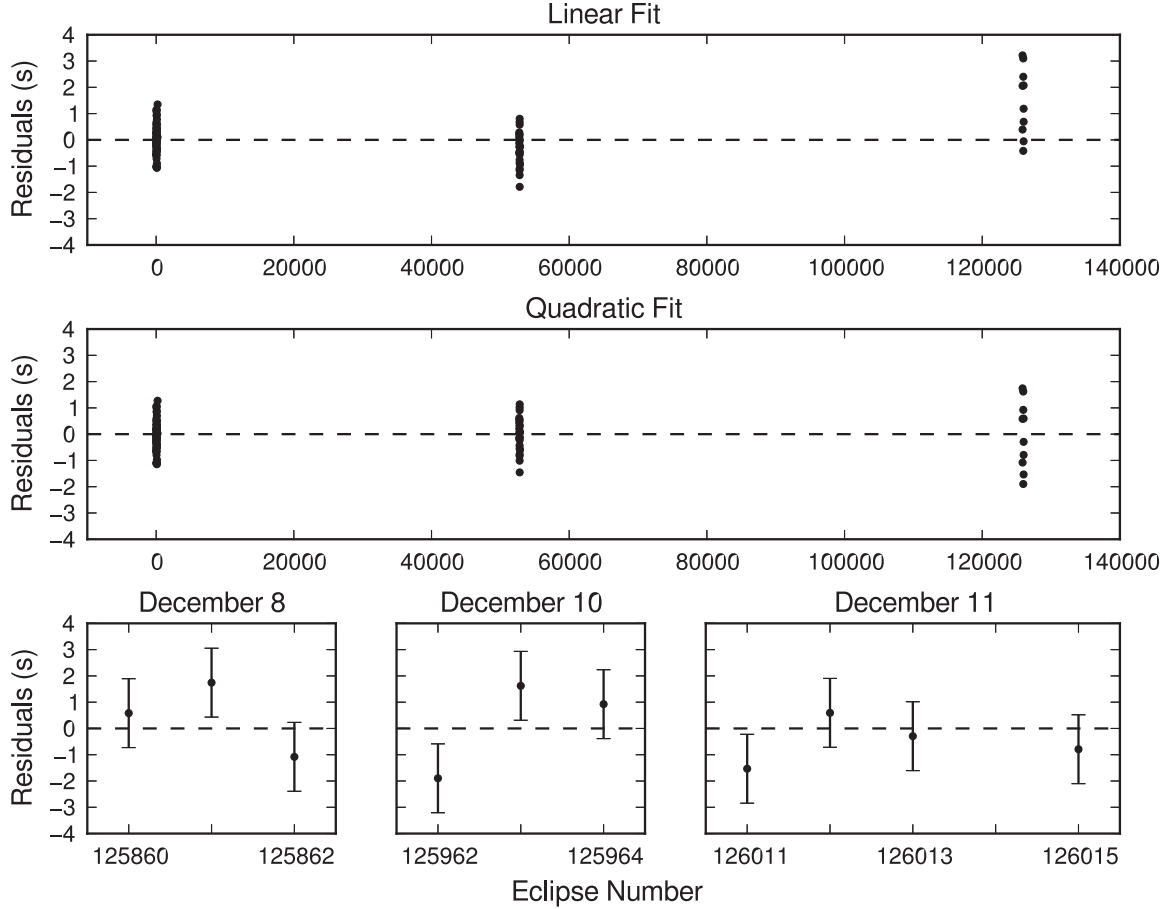}
\end{center}
\caption{(Top) Residuals of the linear fit ephemeris for 2006 and 2009 data published in \citealt{copperwheat11} and the 2012 ARCONS observations. (Middle) Residuals of the quadratic fit ephemeris, for the same data sets.  (Bottom) Plot of the residuals of the quadratic fit ephemeris for the ARCONS data only.
}
\label{fig:ephemeris}
\end{figure*}

Timing errors for the eclipse fits for each epoch (2006, 2009, 2012) are determined by taking all the data points from that epoch, fitting and subtracting a locally determined linear ephemeris, and then taking the standard deviation of these residuals.  This approach appears to be more robust than propagating the photometric errors through the eclipse fitting model as there is some intrinsic variation in the eclipse timing due to flickering in the accretion disk (\citealt{copperwheat11}).

We then combined the eclipse timings calculated from our simple model fit with previous results from \citealt{copperwheat11} to fit a new ephemeris. Again, we used the Levenberg-Marquardt method to determine the fits. There were two models that we used to fit the ephemeris: a linear model with a constant period and a quadratic model with a $\dot{P}$ component. We Taylor expanded the eclipse number in terms of the eclipse time in the form $ N = N_{0} + \nu (t - t_{0})$, where $N$ is the eclipse number, $t$ is the Barycentric Dynamical Time (TDB), in Modified Julian Days (MJD), and $\nu$ is the eclipse frequency. $N_{0}$ is the fit eclipse number of our first eclipse measured at $t_{0} = 56270.513365$ days. We found that the linear ephemeris followed the relation
\begin{align}
N = 125860.0012 + \frac{50.86140661}{days} (t-t_{0}).
\end{align}

\noindent
The measured period with this model is 0.01966127299 $\pm$ 3.0$\times 10^{-11}$ days.

In the qudratic ephemeris model, we added the second order term $\frac{1}{2} \dot{\nu} (t-t_{0})^{2}$, where $\dot{\nu}$ is the frequency time derivative. The quadratic ephemeris followed the relation
\begin{align}
\nonumber
N = &125860.0003 + \frac{50.86140529}{days} (t-t_{0})\\& - \frac{1}{2} \times \frac{7.95 \times 10^{-10}}{days^{2}} (t-t_{0})^{2}.
\end{align}

\noindent
With this model, the measured period at the time of our first eclipse is 0.01966127350 $\pm$ 9.7$\times 10^{-11}$ days. The measured period derivative term, $\dot{P}$, is $(3.07 \pm 0.56) \times 10^{-13}$. This is in range of the anticipated period change of $\dot{P} \sim 3 \times 10^{-13}$ given by \citealt{anderson05}. From our measured $\dot{P}$ and the primary and donor mass values given by \citealt{copperwheat11}, we predict a conservative mass transfer rate, accounting for angular momentum loss from gravitational wave radiation, of $\sim 1.8 \times 10^{-10} M_{\odot}$/yr. This transfer rate is reasonable for AM CVn systems along either the white dwarf or helium star donor formation paths, as shown in Figure 1 of \citealt{nelemans05}.

Plots showing the residuals of the linear and quadratic fits for all of the data as well as more detailed plots of the residuals for only the recently taken data are shown in Figure~\ref{fig:ephemeris}. In the December 11 data, eclipse number 126014 was omitted. Data quality was poor during this time (as can be seen in the bottom panel of Figure~\ref{fig:lightcurve}, 4th eclipse), and the eclipse time could not be measured accurately.

We tested the likelihood of a quadratic ephemeris as opposed to a linear ephemeris. To do this, we measured the goodness of fits by calculating the reduced $\chi^{2}$ values for both models. The linear fit had a $\chi^{2}$ value of 1.24, whereas the quadratic fit had a $\chi^{2}$ value of 1.01. This shows that the quadratic model fits the observational data much better than the linear model. With this we claim a detection of $\dot{P}$ at 5.4 $\sigma$.

\subsection{Quasi-periodic Oscillation Search}

\begin{figure}
\centering
\begin{center}
\includegraphics[width=\linewidth]{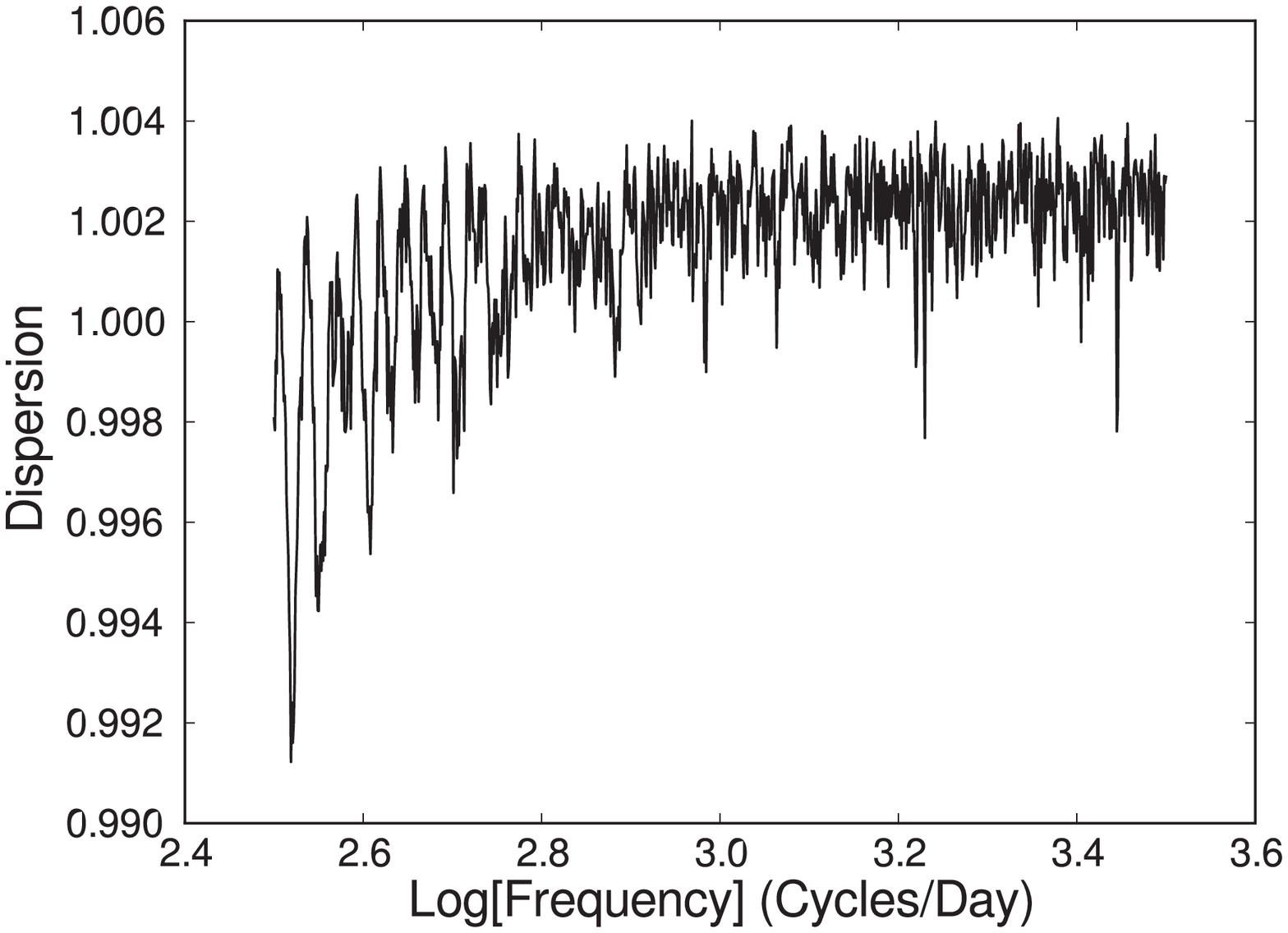}
\end{center}
\caption{Phase dispersion measures of the December 8 data in an intermediate frequency range. The possible QPO observed in the 2006 data by \citealt{copperwheat11} was not seen in our data.
} 
\label{fig:intermediate}
\begin{center}
\includegraphics[width=\linewidth]{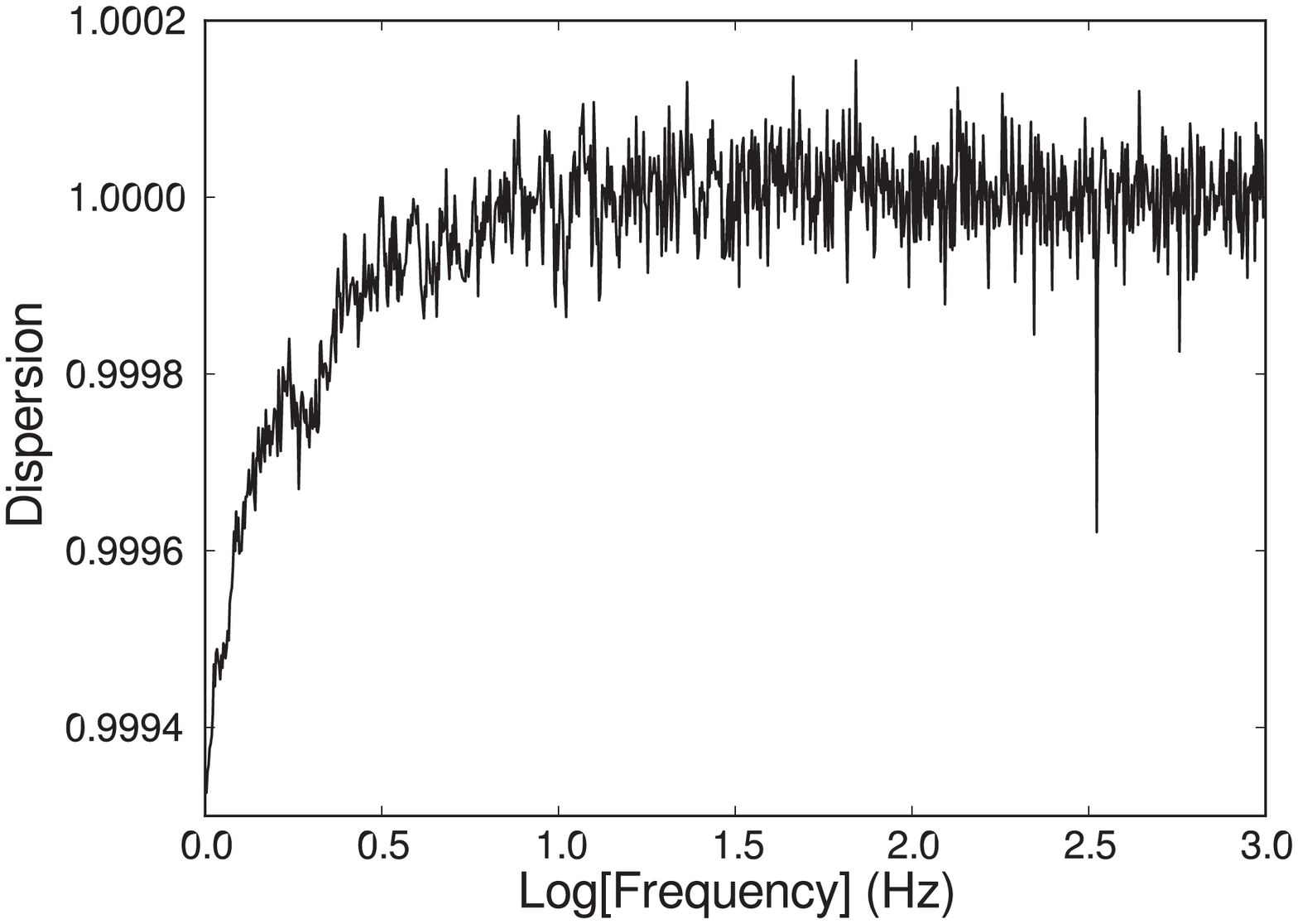}
\end{center}
\caption{Average of many high frequency ($\ge$1 Hz) phase dispersion measures taken of December 8 data. No high frequency QPOs were visible.
} 
\label{fig:high_freq}
\end{figure}

We used the microsecond timing resolution of ARCONS to look for quasi-periodic oscillations (QPOs) in a large frequency range. We first looked at a low-intermediate frequency range ($10^{2.5}$ -- $10^{3.5}$ cycles/day). To do this, we used a method similar to the one used to create the low frequency phase dispersion plots seen in Figure~\ref{fig:pdm}. This involved PSF fitting images that have been integrated for 1s. We then performed PDM using the image times and the flux calculated from the fit parameters. A $10^{2.5}$ -- $10^{3.5}$ cycles/day phase dispersion plot of the December 8 data is shown in Figure~\ref{fig:intermediate}. This method was used for frequencies of up to $\sim$1 Hz, as higher sampling rates made fitting a PSF increasingly more difficult. No evidence of a QPO was found in this range, including the possible QPO seen at $\sim$1700 cycles/day in the 2006 data of \citealt{copperwheat11}. It is worth noting that this QPO signal was not seen in their 2009 data.

We also performed PDM in the $\ge$1 Hz range. For these higher frequency calculations, we used standard aperture photometry to retrieve photon timestamps to a precision of $\sim$10$\mu$s. PSF fitting fails at such small timescales. The photons were binned together to form count per 500$\mu$s intervals. These count rates were then used to perform PDM in blocks of 1s, and the dispersion measures were averaged together to obtain a single dispersion plot. The high frequency ($1$ -- $10^{3}$ Hz) phase dispersion plot for the December 8 data is shown in Figure~\ref{fig:high_freq}. There was also no evidence of a QPO signal at these high frequencies.

\subsection{Spectral Variability}

We examine the variability of SDSS J0926+3624 during various phases of the orbit between four wavelength bands: 4000-5500 \AA, 5500-7000 \AA, 7000-8500 \AA, and 8500-10000 \AA. To do this, we determine the wavelength of an individual photon and place the photon into a corresponding wavelength bin. Finding a PSF fit for the lower energy photons was difficult because the brightness of SDSS J0926+3624 is comparable to the sky background at these energies. Therefore, we used standard aperture photometry in each of the four wavelength bins. 

\begin{figure*}
\centering
\begin{center}
\includegraphics[width=\textwidth]{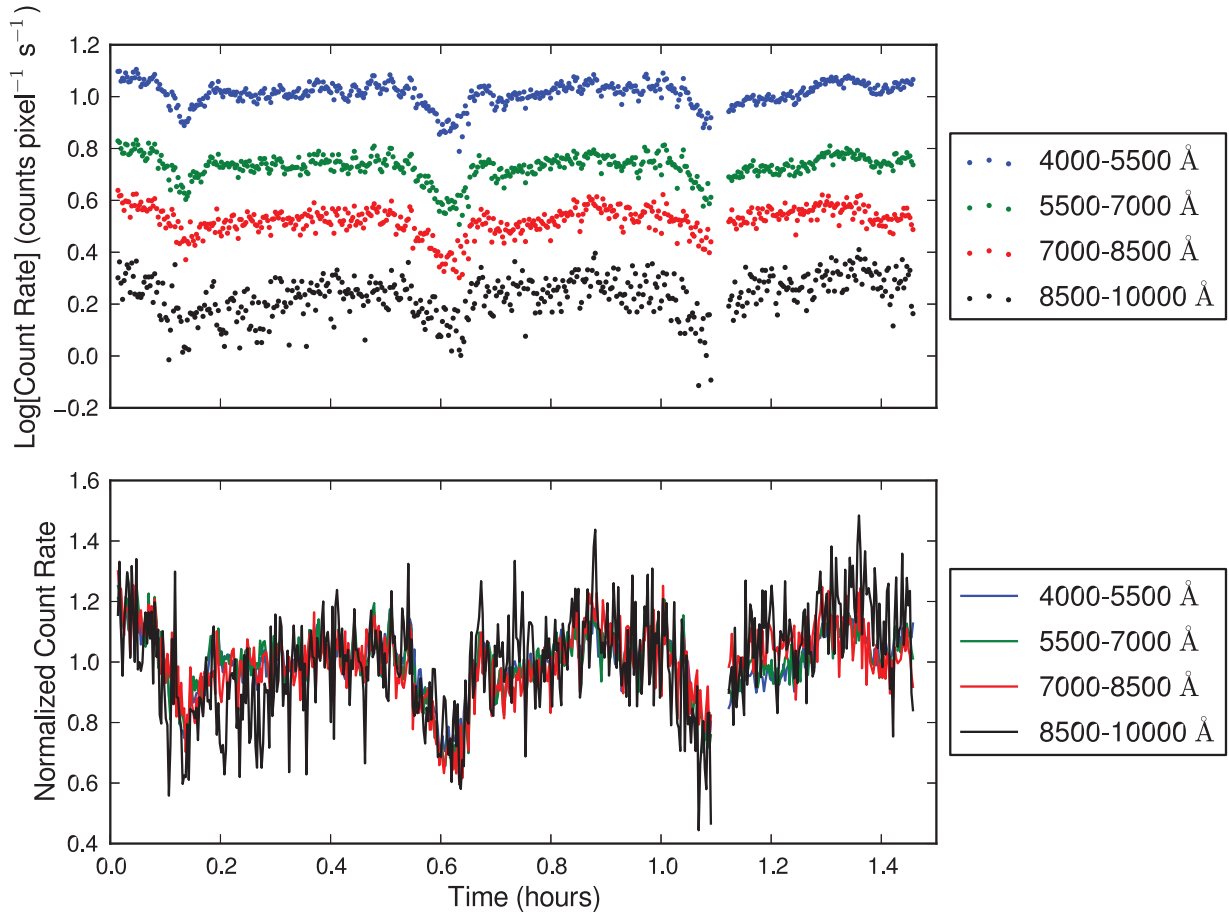}
\end{center}
\caption{(Top) December 8 light curves in the blue (4000-5500 \AA), green (5500-7000 \AA), red (7000-8500 \AA), and infrared (8500-10000 \AA) bands, on a log scale. (Bottom) Light curves are scaled by the mean count rate in each band, in order to show the similarity in the four bands.
}
\label{fig:spectral}
\end{figure*}

In Figure~\ref{fig:spectral}, we plot the resulting light curves for the December 8 data. As expected, the blue (4000-5500 \AA) and green (5500-7000 \AA) bands received much higher count rates than the red (7000-8500 \AA) and the infrared (8500-10000 \AA) bands. Scaling by the mean in each band shows that the light curves were fairly consistent between different bands. There was little spectral variability observed during an eclipse.

\section{Conclusions}
The energy and timing information for individual photons obtainable with current MKIDs allowed us to explore the parameter space of SDSS J0926+3624 in exciting new ways. We were able to study the time variability of SDSS J0926+3624 at higher frequencies (up to 1,000 Hz) than had previously been done, and we showed that no QPOs existed at these frequencies. With the energy information, we showed that there is little spectral variability throughout an orbit. Most importantly, we were able to use the eclipse timing information of our observations to further constrain the orbital period of the system, and we found a $\dot{P}$ of $3.07 \times 10^{-13}$ at a 5.4 $\sigma$ level, consistent with predictions for this system. Observations of SDSS J0926+3624 are planned with updated generations of MKID arrays over the next few years that will improve the SNR of both the aperture photometry and PSF fitting and improve the spectral variability analysis. 

\section*{Acknowledgments}
The MKID detectors used in this work were developed under NASA grant NNX11AD55G, and the readout was partially developed under NASA grant NNX10AF58G. S.R. Meeker
was supported by a NASA Office of the Chief Technologist's Space Technology Research Fellowship, NASA grant NNX11AN29H. This work was partially supported by the Keck Institute for Space Studies. Fermilab is operated by Fermi Research Alliance, LLC under Contract No. De-AC02-07CH11359 with the United States Department of Energy. The authors would like to thank Shri Kulkarni, Director of the Caltech Optical Observatories for facilitating this project, as well as the excellent staff of the Palomar Observatory. This project also greatly benefitted from the support of Mike Werner, Paul Goldsmith, and Jonas Zmuidzinas at JPL.

\bibliographystyle{mn2e}
\bibliography{J0926}

\end{document}